# Hydrogen patterning of $Ga_{1-x}Mn_xAs$ for planar spintronics

*R. Farshchi[a,b], P. D. Ashby[c], D. J. Hwang[a], C. P. Grigoropoulos[a,b], R.V. Chopdekar[a], Y. Suzuki[a], and O.D. Dubon[a,b]

[a]University of California, Berkeley, CA 94720, USA
[b]Lawrence Berkeley National Laboratory, Berkeley, CA 94720, USA
[c]Molecular Foundry, Lawrence Berkeley National Laboratory, Berkeley, CA 94720, USA

**ABSTRACT**

We demonstrate two patterning techniques based on hydrogen passivation of $Ga_{1-x}Mn_xAs$ to produce isolated ferromagnetically active regions embedded uniformly in a paramagnetic, insulating host. The first method consists of selective hydrogenation of $Ga_{1-x}Mn_xAs$ by lithographic masking. Magnetotransport measurements of Hall-bars made in this manner display the characteristic properties of the hole-mediated ferromagnetic phase, which result from good pattern isolation. Arrays of $Ga_{1-x}Mn_xAs$ dots as small as 250 nm across have been realized by this process. The second process consists of blanket hydrogenation of $Ga_{1-x}Mn_xAs$ followed by local reactivation using confined low-power pulsed-laser annealing. Conductance imaging reveals local electrical reactivation of micrometer-sized regions that accompanies the restoration of ferromagnetism. The spatial resolution achievable with this method can potentially reach <100 nm by employing near-field laser processing. The high spatial resolution attainable by hydrogenation patterning enables the development of systems with novel functionalities such as lateral spin-injection as well as the exploration of magnetization dynamics in individual and coupled structures made from this novel class of semiconductors.

*Corresponding Author: email: rfarshchi@berkeley.edu , fax: 510-486-5530
postal address: 210 HMMB, UC Berkeley, Berkeley, CA 94720

The fairly young field of semiconductor spintronics promises to deliver novel device functionalities by harnessing the spin properties of charge carriers [1]. Ferromagnetic semiconductors are expected to play a key role in spintronics due to their efficient spin-injection into non-magnetic semiconductors as well as their intertwined electronic and magnetic properties that lead to external gating capabilities [2, 3]. $Ga_{1-x}Mn_xAs$ is the most well studied ferromagnetic semiconductor [1, 2, 4] in which Mn ions occupy a few atomic percent of the cation sublattice and act as acceptors, providing holes that mediate ferromagnetic exchange between the Mn atoms.

Herein we present techniques to produce planar, isolated, ferromagnetically active regions embedded uniformly and with high spatial resolution in a non-active semiconductor matrix by hydrogenation patterning of $Ga_{1-x}Mn_xAs$. These techniques can be used to realize composite structures featuring ferromagnetic and non-ferromagnetic phases in a structurally identical semiconductor host. The control of ferromagnetism in a semiconductor with high spatial resolution across a surface gives rise to new paradigms in device design based on planar functionalities including lateral spin-injection and gating. Moreover, our methods allow for the development of novel systems for the exploration of magnetic phenomena such as domain wall formation, thermal effects, and magnetic coupling between planar, isolated, ferromagnetic elements.

Hydrogenation of $Ga_{1-x}Mn_xAs$ has the well-known effect of passivating Mn acceptors [5-8]. Hydrogen forms a complex with Mn and a neighboring As atom in which Mn is no longer an electrically active acceptor and hence ceases to provide a hole to mediate inter-Mn ferromagnetic exchange in the film. The effects of hydrogen passivation can be reversed via extended furnace annealing, which breaks up the Mn-H-As complexes and reactivates the Mn acceptors. We demonstrate techniques that utilize *local* hydrogen passivation and depassivation to realize isolated, sub-micrometer, ferromagnetically and electrically active regions embedded in a passivated substrate. The first approach involves selective hydrogen passivation using lithographic masking while the second involves selective hydrogen depassivation using local low-power pulsed-laser annealing.

Hydrogen passivation was carried out selectively by blocking those regions that were to remain ferromagnetically active with a protective layer, such as $SiO_2$. A ~5 mm x 5 mm $Ga_{1-x}Mn_xAs$ sample prepared by ion implantation followed by pulsed laser melting (II-PLM) with x ~ 0.04 ($T_C$ = 100 K) was coated with ~1 μm of $SiO_2$ deposited at 150 °C via plasma-enhanced chemical vapor deposition (PECVD). Details of the II-PLM growth process are given elsewhere [9-11]. The oxide layer was patterned lithographically into the shape of a Hall-bar and inserted into a direct-exposure, radio-frequency hydrogen plasma (13.56 MHz, 100 W) at 130 °C for six hours. The oxide layer was subsequently removed with HF. Metal contacts for electrical measurements were defined via lift-off of e-beam evaporated Pd /Au (100 Å / 500 Å).

The Hall-bar patterns were imaged using conductance atomic force microscopy (C-AFM), which allows mapping of these patterns by measuring the contact conductance associated with a biased scanning AFM tip. Due to the passivating effect of hydrogen on the electrical activity of Mn, a large contrast in contact conductance is observed between active and hydrogenated regions, as shown in Figure 1 for a section of the Hall-bar. The surface roughness was found to be uniform within a few nanometers for these patterns.

Magnetotransport measurements performed on one such Hall-bar are shown in Figure 2. They display the characteristic features associated with the hole-mediated ferromagnetic phase [12] including a strong anomalous Hall resistance (Figure 2a) as well as a reversal of curvature of the magnetoresistance upon cooling below the $T_C$ of 100 K (Figure 2b). The temperature dependence of the sheet resistance, shown in Figure 2c, displays the expected peak at $T_C$ due to increased scattering during the paramagnetic to ferromagnetic phase transition. These characteristics are testament to good pattern isolation; current injected into a poorly isolated Hall-bar would not be confined to the pattern, resulting in very different magnetotransport results.

The active ferromagnetic patterns can be made in the sub-micrometer regime by using suitably sized $SiO_2$ features to block the hydrogenation. For this purpose, a ~200 nm $SiO_2$ layer was deposited via e-beam evaporation through a stencil mask placed in contact with a ~5 mm x 5 mm $Ga_{0.96}Mn_{0.04}As$ sample. The stencil mask contains arrays of square windows defined by e-beam lithography, with the smallest windows being 250 nm on a side. The sample was then exposed to the same hydrogenation conditions mentioned above for three hours followed by oxide removal with HF. C-AFM images for arrays of electrically active $Ga_{1-x}Mn_xAs$ features prepared in this fashion are shown in Figure 3. The magnetic properties of these features are expected to be similar to those of the Hall-bar patterns described above as the same hydrogenation conditions have been used.

An alternative technique for creating isolated $Ga_{1-x}Mn_xAs$ features in a passivated substrate is selective laser depassivation. It is known that the hydrogen ions can be dissociated from the Mn atoms via extended furnace annealing at ~190 °C [6], and consequently the magnetic and transport properties of the as-grown film can be restored. To produce isolated active regions, it is possible to mimic the effect of furnace annealing *locally* using low-power pulsed-laser annealing with a confined spot-size to reactivate desired locations on the hydrogenated substrate. For this purpose, $Ga_{0.96}Mn_{0.04}As$ samples prepared via II-PLM were hydrogenated repeating the above conditions for six hours. Laser annealing was then carried out using a Q-switched Nd:YAG laser ($\lambda$ = 532 nm, FWHM = 4-6 ns) at 10 Hz for 5 minutes. The laser spot was focused with a 0.55 NA (50X) objective lens to a diameter of ~ 4μm. C-AFM measurements for a feature made with a fluence of ~65 mJ/cm$^2$ are shown in Figure 4. A large contrast in contact conductance is accompanied by a highly uniform topography, suggesting that under these conditions the laser energy is sufficient to remove hydrogen but not sufficient to induce significant structural damage or ablation. To investigate the degree of magnetic reactivation, laser annealing was carried out at a fluence of 65 mJ/cm$^2$ using a 3mm x 3mm homogenized beam (with otherwise similar laser conditions), allowing for measurement with SQUID magnetometry. Partial Mn depassivation was observed, reflected by a 60% recovery of the Curie temperature to $T_C$ = 60 K. The attainable spatial resolution of this method can potentially reach <100 nm by performing laser depassivation under near-field configurations.

In summary, hydrogenation patterning has been used to realize isolated, ferromagnetically active $Ga_{1-x}Mn_xAs$ regions embedded in a non-active host. The planar integration of impedance matched, ferromagnetic elements creates unique design possibilities for lateral spin-injection and gate-modulated spintronic devices. While hydrogenation patterning has been demonstrated in the $Ga_{1-x}Mn_xAs$ system, it can be extended to other ferromagnetic semiconductors including $Ga_{1-x}Mn_xP$, which displays a similar hydrogen passivation behavior.

The authors would like to acknowledge A. Liddle and J. T. Robinson for stencil mask fabrication and design, N. Misra for assistance in laser processing, and I. D. Sharp, J, W, Beeman, and E. E. Haller for ion-implantation. This work was supported in part by the National Science Foundation under contract number DMR-0349257 and in part by the Director, Office of Science, Office of Basic Energy Sciences, Division of Materials Sciences and Engineering, of the U.S. Department of Energy under Contract No. DE-AC02-05CH11231. R. Farshchi acknowledges support from an Intel Fellowship.

**FIGURE CAPTIONS:**

Figure 1: C-AFM image of Hall-bar produced via selective hydrogenation (3 hours). Inset shows microscope image of protective oxide pattern after hydrogenation.

Figure 2: Magnetotransport measurements on a $Ga_{1-x}Mn_xAs$ Hall-bar produced via selective hydrogenation (6 hours): a) field dependence of anomalous Hall resistance, b) field dependence of sheet resistance, c) temperature dependence of sheet resistance.

Figure 3: C-AFM image of sub-micron $Ga_{1-x}Mn_xAs$ features produced with selective hydrogenation (3 hours). Inset shows a scan on a single feature.

Figure 4: C-AFM height (top) and conductance (bottom) images of a feature resulting from low-power pulsed-laser annealing of hydrogenated $Ga_{1-x}Mn_xAs$ (6 hours hydrogenation).

**FIGURES:**

Figure 1:

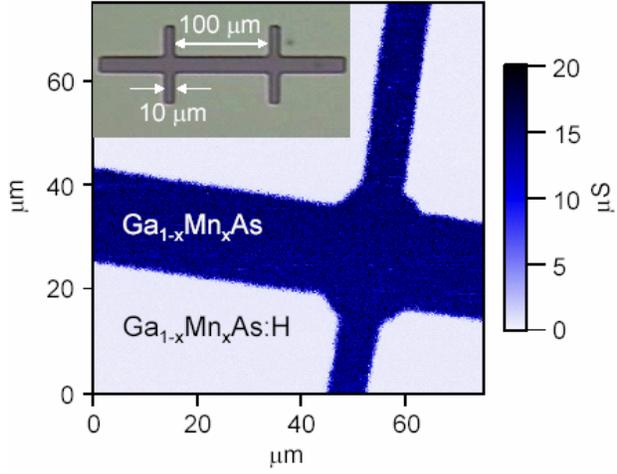

Figure 2:

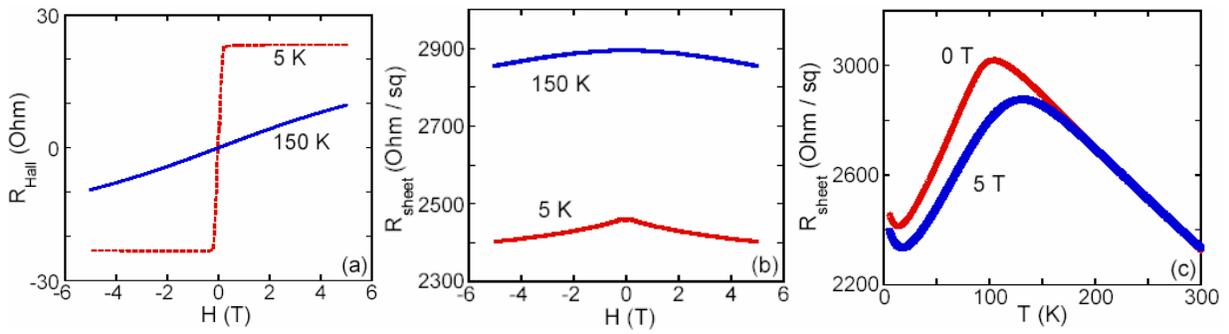

Figure 3:

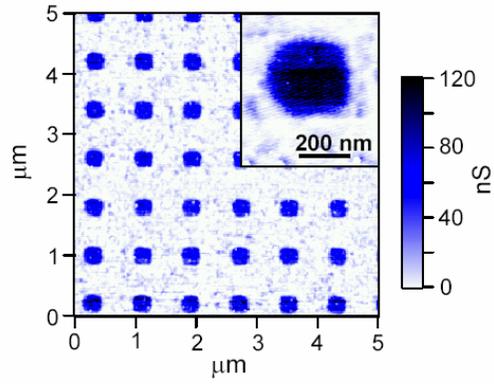

Figure 4:

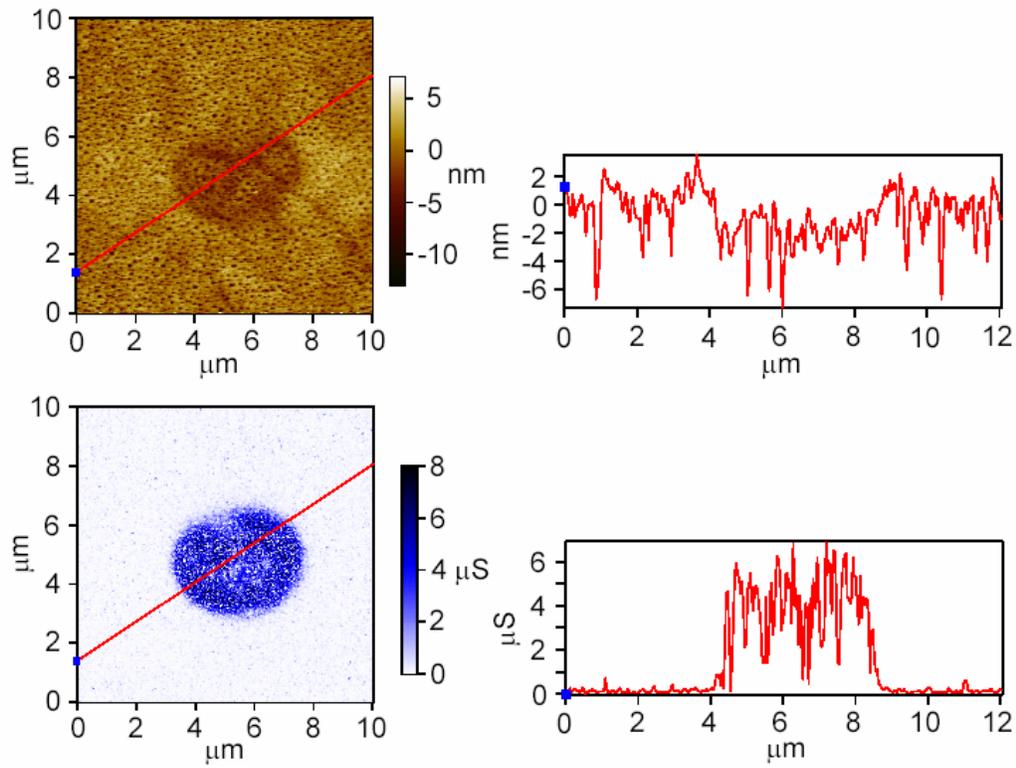